\documentclass{article}

\usepackage{PRIMEarxiv}

\usepackage[utf8]{inputenc} 
\usepackage[T1]{fontenc}    
\usepackage{hyperref}       
\usepackage{url}            
\usepackage{booktabs}       
\usepackage{amsfonts}       
\usepackage{nicefrac}       
\usepackage{microtype}      
\usepackage{lipsum}
\usepackage{fancyhdr}       
\usepackage{graphicx}       
\usepackage[numbers]{natbib}
\usepackage{amsmath}
\usepackage{enumitem}
\usepackage{algorithm}
\usepackage{algpseudocode}
\graphicspath{{media/}}     

\title{A network simulation of OTC markets with multiple agents}

\author{
  James T. Wilkinson \\
  McCormick School of Engineering \\
  Northwestern University \\
  Evanston IL, USA\\
  \texttt{jameswilkinson2022@u.northwestern.edu} \\
  \And
  Jacob Kelter \\
  Computer Science and Learning Sciences \\
  Northwestern University \\ 
  Evanston IL, USA \\
  \texttt{jacobkelter@u.northwestern.edu} \\
  \And
  John Chen \\
  Computer Science and Learning Sciences \\
  Northwestern University \\ 
  Evanston IL, USA \\
  \texttt{yuehanchen2023@u.northwestern.edu} \\
  \And
  Uri Wilensky \\
  Computer Science and Learning Sciences \\
  Northwestern University \\ 
  Evanston IL, USA \\
  \texttt{uri@northwestern.edu}
}

\begin{document}
\maketitle



\begin{abstract}
We present a novel agent-based approach to simulating an over-the-counter (OTC) financial market in which trades are intermediated solely by market makers and agent visibility is constrained to a network topology. Dynamics, such as changes in price, result from agent-level interactions that ubiquitously occur via market maker agents acting as liquidity providers. Two additional agents are considered: \textit{trend investors} use a deep convolutional neural network paired with a deep Q-learning framework to inform trading decisions by analysing price history; and \textit{value investors} use a static price-target to determine their trade directions and sizes. We demonstrate that our novel inclusion of a network topology with market makers facilitates explorations into various market structures. First, we present the model and an overview of its mechanics. Second, we validate our findings via comparison to the real-world: we demonstrate a fat-tailed distribution of price changes, auto-correlated volatility, a skew negatively correlated to market maker positioning, predictable price-history patterns and more. Finally, we demonstrate that our network-based model can lend insights into the effect of market-structure on price-action. For example, we show that markets with sparsely connected intermediaries can have a critical point of fragmentation, beyond which the market forms distinct clusters and arbitrage becomes rapidly possible between the prices of different market makers. A discussion is provided on future work that would be beneficial.
\end{abstract}





\section{Introduction}

Agent-based modelling (ABM) has become a popular method for simulating financial markets since its initial use in the late 1980s \cite{takahashiJASSS, kim_markowitz_1989, palmer_brian, samanidou_zschischang_stauffer_lux_2007, levy_2008}. Financial risk modelling intuitively lends itself to this approach. Firstly, real world markets are naturally defined by the interaction of multiple agents, each with a diverse set of functions and behaviours. Secondly, systems with many interacting sub-units often exhibit scale-free phenomena, as is the fractal nature of market prices \citep{mandelbrot_fisher_calvet_1997}. Emergent phenomena within markets are often difficult to model analytically, and whilst such models exist \citep{hagan2014arbitrage, hagansmile, bartlett2006hedging}, they are inherently limited by their high-level assumptions of a predefined stochastic process governing the price movements. The complexity and nature of the financial environment has encouraged the use of ABM, which typically makes fewer assumptions about pricing dynamics. Now a staple component of financial risk analysis \citep{hassan2018survey}, the multi-agent approach allows a better understanding of how certain macroscopic phenomena emerge from microscopic trader behaviours.

Prior works into agent-based financial modelling have used a variety of agents and mechanisms to define the system's behavior \citep{vanfossan_dagli_kwasa_2020, palmer_brian, amine, samanidou_zschischang_stauffer_lux_2007, levy_2008, kizaki_saito_takahashi_2021, grothmann, toth_scalas_huber_kirchler_2007, johnson_lamper_jefferies_hart_howison_2001, preis_golke_paul_schneider_2006, jefferies_hart_hui_johnson_2001, thurner_farmer_geanakoplos_2010}. Often, these models are specialised for specific investigations, such as exploring the effect of insurance methods on market crashes \citep{kim_markowitz_1989} or increased leverage on the kurtosis of the resulting price distribution \citep{thurner_farmer_geanakoplos_2010}. Substantially all investigations have neglected over-the-counter (OTC) systems in favor of exchange traded markets or auctions \citep{preis_golke_paul_schneider_2006, toth_scalas_huber_kirchler_2007, jefferies_hart_hui_johnson_2001}, and although some of these studies have considered market makers \citep{kirilenko2017flash, paddrik2017effects, paddrik2012agent, bookstaber2015agent}, they are considered symmetrically to the other investor classes and do not behave directly as an intermediary, typically trading through a limit-order book via an exchange. This is partly because the additional requirement to model market maker behaviour, whose reactions to trades can be difficult to quantify. However, market makers play a vital role in the financial system by providing additional liquidity to investors. Their function is especially central to OTC markets in products where no exchange exists. In addition, prior works have never considered a financial market with limited visibility between participants. This is another important factor that applies most relevantly to OTC markets, in which not every trade that occurs is visible to all of the participants. 

We present a new, open-source agent-based model that simulates price changes within a financial market outside of the exchange setting, whilst also incorporating a parameterized visibility between investors. The model uses three types of agents distributed in a network: value investors, trend investors and market makers. Value investors are initialised with price targets, such that they will seek to buy the product if its current price is below the price target, and vice versa if its current price is higher. Trend investors represent those that rely on technical analysis, using a deep Q-learning approach \citep{mnih2013playing} to inform investment decisions by analysing the product's visible price history. Trend investors incorporate a convolutional neural network in this process. Market makers intermediate all of the trades, offering to take the opposite end but at a price of their choosing. Market makers are in direct competition with each-other as trend and value investors always select the most competitive market maker to trade with. In order to keep their prices as accurate as possible, market makers adjust their prices based on the trades they see happening, ensuring that their mid-price reflects the current clearing level by always moving it to the last visible trade price. In order to reflect the market makers' goal to recycle risk, they will additionally move their prices higher or lower depending on their inventory. We also incorporate a soft risk limit to market makers, which has been shown to be an important factor in financial markets, and has empirically been linked to flash crashes \citep{kirilenko2017flash}.

In this paper, we explore the ability of the proposed model to reproduce emergent phenomena observed in real financial markets. In particular, we demonstrate that the model produces fat-tailed distributions of price changes that follow a power-law rank distribution \citep{kim_yoon_chang_2004, warusawitharana_2016}; that prices converge to the mean views of value investors; and that the model exhibits realistic, reproducible and predictable features within its price patterns (often referred to as ``technicals"). In addition, we use the network feature to demonstrate that our market exhibits higher kurtosis when distributed across increasingly sparse networks, and also a critical point of fragmentation, beyond which arbitrage between the prices of different market makers becomes possible.

\section{The Model}

We use the NetLogo platform \citep{wilensky1} to simulate our agent-based model. NetLogo is a widely used platform for agent-based modelling, and has been used to study complex group dynamics from flocking to predator-prey interactions \cite{wilensky4, wilensky5}. NetLogo provides an intuitive interface that allows users to adjust parameters, run experiments and visualise the phenomena being modelled, and as such, is perfect for an open-sourced model such as the one presented in this paper.

The model comprises three types of agent: market makers, value investors and trend investors distributed on a network topology. The behaviours of market makers and value investors are pre-determined, whilst that of trend investors is determined using the machine learning algorithm \textit{deep Q-learning} \citep{mnih2013playing, mnih_human-level_2015}. At each time step, an investor (which comprises a value investor, a trend investor, or a market maker in breach of their soft limit) is selected at random, and allowed to execute a trade versus a market maker that is connected via the network. Once the trade has occurred, the agent's inventory is updated, along with that of the market maker acting as counter party. At each time step, all market makers are allowed to update their prices. The exact mechanism will be laid out in full.

\subsection{Network formation}
\label{sec:network}

\begin{figure}
    \centering
    \includegraphics[scale=0.45]{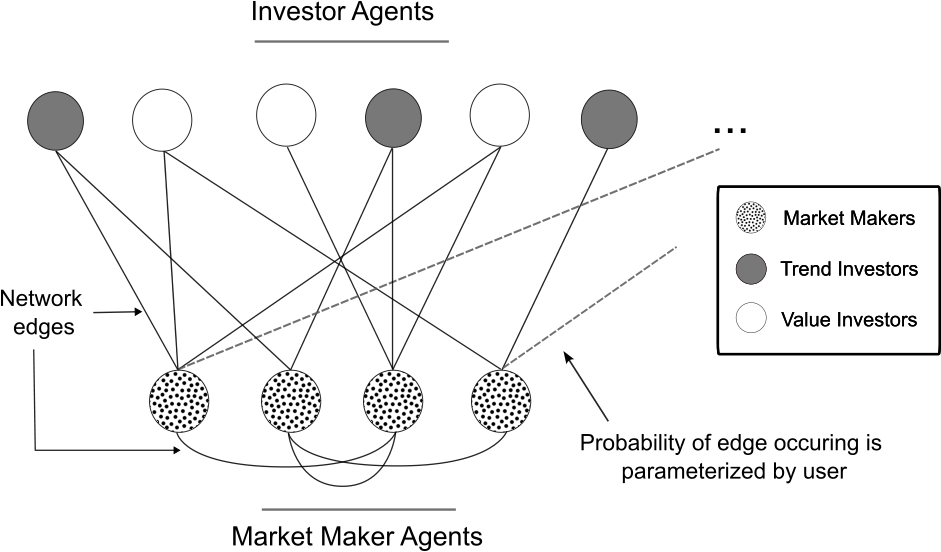}
    \caption{An illustration of our network generation. All possible edges are considered between every agent, with every market maker. Each possible edge is formed with probability $p$, which is a user-defined parameter. Market makers are also interconnected by this same network. Every edge within the full network must be connected to at least one market maker. This aligns with our goal of having all trades performed versus an intermediary market maker.}
    \label{fig:network-build}
\end{figure}

Within the model, interaction and visibility of all agents is constrained to occur along the edges of a network. We formalise this using the network model outlined in Figure \ref{fig:network-build}.

Reflecting that all interactions within our model must occur through market makers, we enforce that at least one market maker must be involved in each network edge, which is a constraint that deviates our network topology from that of a typical random network. The process outlined in figure \ref{fig:network-build} connects each agent to each market maker with a user-defined probability $p$ that also parameterizes the network. Market makers are also connected to one another as part of this process, but no self-connections are allowed. We also enforce that at minimum, each agent is connected to at least one other agent that is a market maker to mitigate the possibility of an agent being entirely isolated from the system.

The network defines the visibility of each agent with the goal of inhibiting the transfer of information throughout the market. As such, within our model, market makers will only react to trades that occur with a market maker that is a direct network neighbor. This visibility in the real world would be representative of price discovery through the inter-dealer broker market. Similarly, trend investors that analyse the price-history only consider those prices that were advertised by market makers that are direct neighbors on the network. 

We notice that the network architecture implied by a system with a central limit order book, as has been used in the majority of the prior literature, resembles a star network, with all agents connected to one single intermediator. Our inclusion of multiple intermediators (market maker agents) drastically increases the complexity within our model, allowing us to perform investigations into the effect of market structure on pricing dynamics that would be beyond the capability of a limit order book model. We perform such investigations ourselves in section \ref{sec:networkresults}.

\subsection{Market Makers}

Market makers always advertise two prices - a bid price at which they are willing to buy, and an offer price at which they are willing to sell. The difference between these prices is referred to as the bid-offer, and is a fixed parameter determined by the user that is constant across all market makers within our system. We refer to the middle of these prices as the \textit{mid} price. Defining a market maker's mid price as $P_{mid}$, with bid-offer $b$, a market maker's price when either buying or selling is given by equation \ref{eq:mmpx1}.

\begin{equation}
    P^{m} = P_{mid}
       + f \frac{b}{2}
    \label{eq:mmpx1}
\end{equation}

where

\begin{equation}
    f = 
    \begin{cases}
        $+1$ & \text{if market maker is selling}\\
        $-1$ & \text{if market maker is buying}
    \end{cases}
\end{equation}

Equation \ref{eq:mmpx1} formalises that a market maker's buy price is lower than their sell price. Market makers calculate their mid by considering both the last trade price that they observed (through a neighboring market maker on the network) and their inventory. Parameterizing the sensitivity of market maker prices to their inventory level with the proportionality $a$, we define a market maker's mid price as in equation \ref{eq:mmpx4}.

\begin{equation}
    P^m = P^{m}_{last}
       + f \frac{b}{2}
       - a I^m
       \label{eq:mmpx4}
\end{equation}

where $P^{m}_{last}$ is the last trade price seen by market maker $m$, and $I^m$ is their current inventory level. 


The simulation begins by setting an initial price of 100, and market makers initialise their mid to this level. On the first iteration, all market makers begin by advertising the same prices, and the first transacting agent will choose their counter-party randomly from the symmetrical market makers. However, the market maker behaviours will diverge as their inventories evolve. When executing a trade, we consider the market maker's inventory \textit{following} the transaction, so that if market maker $m$ had an initial inventory $I^m$, and was transacting $S$ units with a bid offer $b$, they propose a transaction price given by $P^m$ in equation \ref{eq:mmpx2}.

\begin{equation}
    P^{m} = P^{m}_{last}
       + f \frac{b}{2} 
       - a \big ( I^m + S \big )
    \label{eq:mmpx2}
\end{equation}

The value of $I^m + S$ is positive if the market maker is net long following the transaction, and negative if the market maker is net short. Our choice of a linear relationship between price and resulting inventory level is arbitrary, and a number of alternative functional forms were also tested (specifically quadratic and exponential forms along with varying values for $a$). However, no significant effect was observed, and so we progress with the functionality outlined in equations \ref{eq:mmpx4} and \ref{eq:mmpx2}, and our value of $a$ being 0.001. This formulation implies that market makers will alter their prices lower when looking to reduce inventory, and higher when looking to increase inventory, which is qualitatively what we desire. The transacting counter party will always select the market maker with the most competitive price. If the most competitive price is shared by two or more market makers, one is chosen at random to trade with.


\subsection{Value investors}

At initialisation, each value investor is given a \textit{target} price of the product, which is drawn from a bimodal Gaussian distribution with a mean of 100. Our consideration of a bimodal distribution allows us to model common market scenarios where investors take views on a binary outcome, as is common around events such as political elections. The mean target price of 100 is arbitrarily chosen, and the user has full control over the distance between the centroid of each Gaussian mode so that a conventional, single-mode Gaussian can be used by setting this distance to zero. The targets of the value investors are static, and never change. Heuristically, this allows us to interpret the value investor target as a long-term view on the price that doesn't change over the time frame of our simulations.


A value investor’s action is to compare the prices of market makers, and choose to buy if the most competitive offer is lower than their target price and to sell if the most competitive bid is higher than their target price. In scenarios where a value investor can profit from both buying and selling versus different market maker prices, the investor will choose the action that maximises the difference between the trade price and their target price, thus maximising their expected profit. As is typical in a real financial market, the trade size is variable, and to reflect value investors' reaction to differing opportunity sizes, we determine that their trade size grows linearly as the difference between the investor’s target and the market price increases. Specifically, value investors calculate their trade size as

\begin{equation}
    S = \frac{P - T_{v}}{\sigma}
    \label{eq:size}
\end{equation}

Where $P$ is the most competitive transaction price provided by the market makers, $T_{v}$ is the value investor's target, and we choose a proportionality of $\sigma=5$ which was found to generate an appropriate distribution of trade sizes involving value investors. The trade size is capped to a maximum value determined by the user. This is an intuitive restriction found within real financial markets. Equations \ref{eq:mmpx2} and \ref{eq:size} are simultaneous equations with an analytic solution for a trade price and size that satisfies both the value investor and the market makers behavior when they interact. Heuristically, this approach formalises the negotiation mechanics that would typically occur between a market maker and their client during a trade. In practice, we use the solution to determine the trade characteristics that consistently satisfy the behaviour of both market makers and value investors whilst transacting. Equation \ref{eq:analyticSolution} presents this solution.

\begin{equation}
    P^m =\bigg( P^{m}_{last} + f \frac{b}{2} - a ( I^m - \frac{T_{v}}{\sigma_{}} ) \bigg) \bigg(1 + \frac{a}{\sigma_{}} \bigg)^{-1}
    \label{eq:analyticSolution}
\end{equation}

Meanwhile, market makers are also allowed to trade versus one another in order to distribute risk between them via the inter-broker market. Each market maker that is beyond their position limit is allowed to act as an investor, asking the other market makers that are also network neighbours to trade a size that would reduce their inventory to zero in a single trade, capped at the maximum allowed trade size defined by the user. The logic here is similar to before, with the transaction price calculated using equation \ref{eq:mmpx2}. This behaviour can be toggled by the user, as outlined in section \ref{sec:params}.


\subsection{Reinforcement Learning Agent Mechanics}
\label{deepQsec}

Recently, interest in the intersection between reinforcement learning (RL) and agent-based modelling has exploded in popularity \citep{tianxu2022applications, canese2021multi, terry2021pettingzoo, vinyals2019grandmaster, wiering2000multi}. RL has offered the means to explore multi-agent environments that require complex and nuanced agent strategies. In turn, the ABM approach has become an integral component of leading RL systems, where it is commonplace to use ABM as part of the training process. These synergies have led to the sub-field of Multi-Agent Reinforcement Learning (MARL) \citep{zhang2021multi}, which find use in the state-of-the-art systems solving autonomous driving \citep{shalev2016safe}, the game of Go \citep{silver2016mastering, silver2017mastering}, and questions in finance \citep{lee2007multiagent}.

We choose to construct our trend investors as artificially intelligent agents, with the goal of simulating investors that use technical analysis to guide their trading decisions. As in the real market, these trend-spotting investors act to remove inefficiencies within the system by profiting from their exploitation. Repeatable price patterns within the system can evolve over time as trend investors dynamically learn to profit from them. Once enough trend investors have recognized this same strategy, their collective investment activity can reduce, and even eradicate the pattern altogether. As such, the inclusion of artificially intelligent trend investors opens the doors to exciting and new research investigating the interplay of dynamically competing RL agents, the temporal evolution of stock price patterns, and the impact that different trading strategies can have on the underlying market.

\begin{figure}
    \centering
    \includegraphics[scale=0.85]{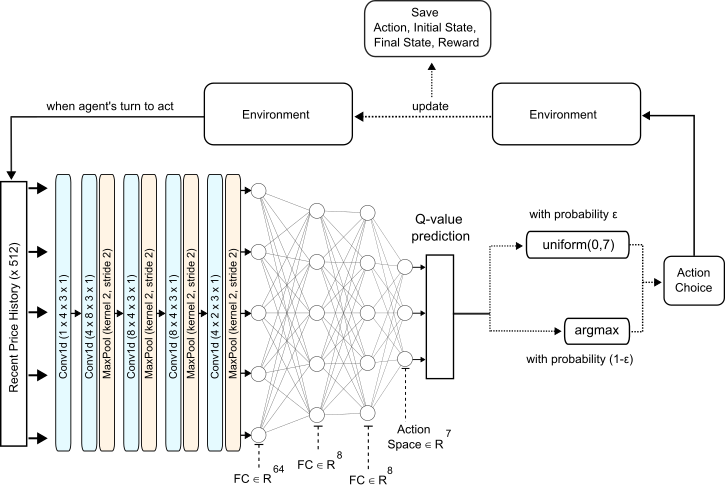}
    \caption{Illustration of our convolutional Q-network in the decision-making process. The dimensions of the convolutional layers are presented in the format (input channels, output channels, kernel size, and stride). All convolutions are one dimensional. When prompted to act, the trend investors either choose to act randomly (with a time-decaying probability $\epsilon$), or to use their Q-network to choose the action that is estimated to maximise a target value $Q$ as defined by Bellman's equation \ref{eq:bellman}. This \textit{epsilon-greedy} approach promotes a healthy mixture of exploration and exploitation during the agent's training.}
    \label{fig:NNimage}
\end{figure}

We use deep Q-learning \citep{mnih2013playing, mnih_human-level_2015} with a convolutional neural network as outlined in figure \ref{fig:NNimage}. Within this framework, our state vector is defined as the most recent 512 historical price data points. Each element within the state vector, for a given trend investor, is formed from the mean of all network-connected market maker mid prices at that snapshot in time. This implies that the state vectors of two trend investors are not necessarily equal, even if they are sampled at the same moment in time. This set-up allows us to define the Q-value associated with taking action $a$ from state $s_t$ using Bellman's equation, outlined in equation \ref{eq:bellman}.

\begin{equation}
    \label{eq:bellman}
    Q(s_t, a) = R(s_{t}, a) + g \max_{a^{\prime}} Q(s_{t+1}, a^{\prime})
\end{equation}

where $R(s_t, a)$ is the reward, which we measure as the profit resulting from executing a trade defined by action $a$, from state $s_t$, $g$ is a discount rate used to attenuate the value of future rewards, and $s_{t+1}$ is the next state vector, which is the state vector for the agent at the moment that a trade is closed and the reward is realised. The central theme to deep Q-learning is to use a neural network to estimate the values of $Q$.

We further simplify the deep-Q paradigm by enforcing that trend investors consider trades individually, and the Q-network aims at choosing the action that maximizes the profit on this particular trade, without regard of the trend investor's other positions. This is a preferable approach to model our trend investors with, because it presents a simpler training paradigm for the model to learn in. In defining the Q value as the profit associated with a trade, plus a discounted optimal value of the resulting state, we achieve consistent training stability, and meaningful trading strategies being adopted by the trend investors.

The Q-network takes the most recent 512 items of the price history as input, which is passed through the convolutional and max-pooling layers and a four-layered fully-connected head with one output for each possible agent action. 

A trend investor can choose between seven actions, when prompted to trade. These seven actions span the full six combinations of trade-directions (buy or sell) and holding-times (100 ticks, 250 ticks or 500 ticks); and also the additional action allowing them to do nothing, which can be useful in the case where all trades are considered unprofitable. All trading actions have the same trade size. This is because variable trade sizes should produce rewards that are direct multiples of one another, such that a well trained agent would always choose the action with the largest allowed trade size. Therefore, we choose to define all trend investor trades to use the maximum trade size allowed within the system, which is a user defined parameter, defaulted at 3 units for this paper's results. The trade time is specified as the number of time steps that must pass before a trend investor's trade can be closed. If the trade-time for a specific trade becomes zero (or negative) during the model's progress, the trend investor will deterministically close this trade at their next turn, at which point the reward (the profit) for the trade is calculated and used to train the agent. At this point, we save four observations in the agent's replay memory: the initial state that was present at the time of the trade, the action chosen, the reward that resulted, and the state vector of the environment when the trade was closed.

Following each action of a given agent, we sample a batch from the agent's memory, and train the Q-network using a mean-squared error loss on equation  \ref{eq:bellman} so that our Q-network learns to predict the value associated with actions given an initial state. Each time this update occurs, the parameter $\epsilon$, which is responsible for varying the agent's proportions of exploration and exploitation, is reduced by a constant factor (we chose to use a factor of $0.995$, which produced stable training).

The chaotic nature of price movements within the model poses a challenge for stable training, especially in the context that deep Q-learning has notorious instability. In the context of our trend investors, unstable training would result in unprofitable and non-meaningful investment strategies. In order to address these stability issues during training, we incorporate a soft-update to the Q-network as first presented by Mnih et. al, whose architecture closely resembles that of our own \citep{mnih_human-level_2015}. For completeness, our algorithm is outlined in \ref{alg:1}.

\begin{algorithm}
\caption{DQN algorithm}
\label{alg:1}

\begin{algorithmic}

\State Randomly initialize Q-network $Q$ 
\State Duplicate Q-network to form the target-network $Q^{\prime}$
\State Initialize replay buffer $R$
\For{episode = 1, E}

\For{t = 1, T}
\State Observe state $s_t$

\State Select optimal action $a_t \gets \underset{a}{\operatorname{argmax}} Q_{i^{\prime}} (s_t, a)$

\State Save the state $s_t$, action $a_t$, observed reward $r_t$ and the next state $s_{t+1}$ into $R$.

\State Sample random minibatch of transitions $(s_i, a_i, r_i, s_{i+1})$ from $R$
\State Set the target value $y_i \gets r_i + \gamma \max_{a} Q^{\prime}(s_{i+1}, a)$
\State Update Q-network using loss: $L \sim \sum_{i} (Q(s_i, a_i) - y_i)^2$

\State Update target network's parameters $\theta^{\prime}$ using the Q-network parameters $\theta^Q$.
\begin{center}
$\theta^{\prime} \gets (1 - \tau) \theta^Q + \tau \theta^{\prime}$
\end{center}

\EndFor
\EndFor

\end{algorithmic}

\end{algorithm}

\begin{figure}
    \centering
    \includegraphics[scale=0.40]{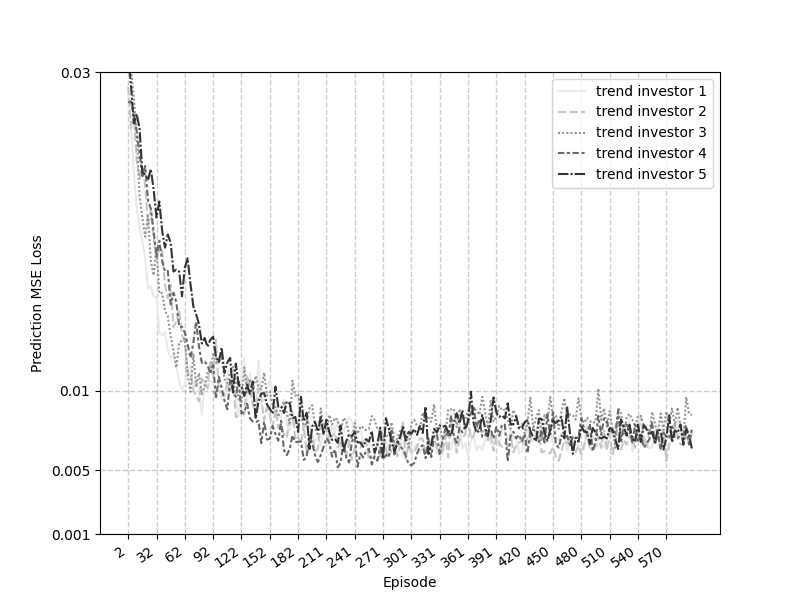}
    \caption{Loss curves demonstrating the convergence of the trend investors' q-networks. The logarithmic y-axis shows the mean-squared error between the model's predicted profit from its trade and the profit that was realised.}
    \label{fig:learning_curves}
\end{figure}

Before generating each set of results, we train the newly-initialised trend investors in the environment until they have reached convergence, as defined by the value of $\epsilon$ reaching a minimum allowed value of $0.05$, and the mean-squared error of the Q-network converging as displayed in figure \ref{fig:learning_curves}. Some prior work \citep{paddrik2017effects} has also introduced an additional set of agents that act randomly in order to capture the effect of additional investor behaviour, and so we note that our trend investors fulfill this same role because of their non-zero terminal value of $\epsilon$, which leaves them infrequently performing random actions as the simulation progresses. 


\begin{figure}
    \centering
    \includegraphics[scale=0.40]{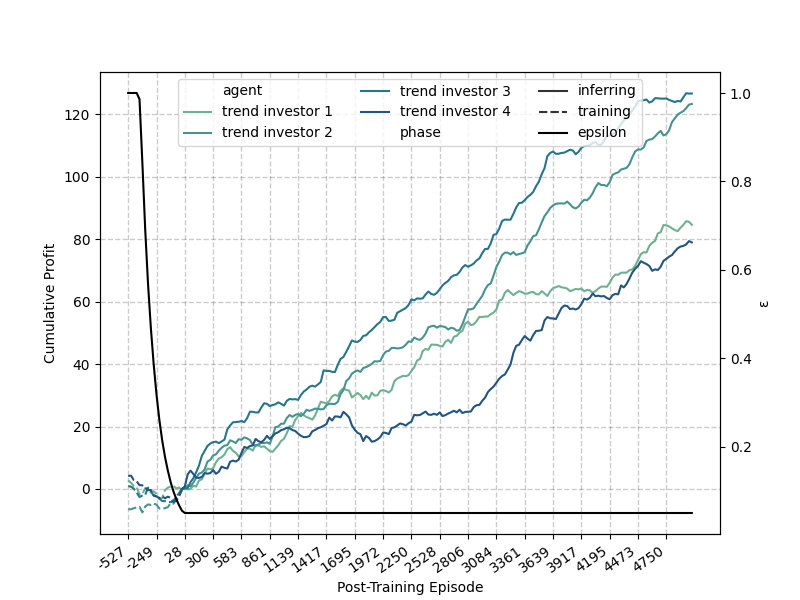}
    \caption{Trend investor profit tracked over a large number of model iterations. Once trained, trend investors are employing profitable and economically viable strategies. Units for the x-axis are model iterations passed following the point at which all trend investor's exploration parameters $\epsilon$ have reached their terminal value of 0.05. As such, negative values along this axis represent time steps where the trend investors are mostly untrained, and their behaviour is mostly exploratory.}
    \label{fig:profit}
\end{figure}

Figure \ref{fig:learning_curves} shows the mean-squared error loss of the Q-network on logarithmic axes, with a decay rate of $\epsilon$ of $0.995$, a learning rate of $1 \times 10^{-3}$ with an Adam optimizer  \citep{kingma2014adam}, and a soft-update parameter $\tau=1e-3$. For the environment, the default parameters outlined in \ref{sec:params} were used. To put an intuition on the scale of the MSE loss shown in figure \ref{fig:learning_curves}, the loss associated with a randomly acting agent trading with trade times of 250 time steps is approximately $27\%$ (price standard deviation is approximately 9\%, and using a maximum trade size of 3), which translates into a standard deviation of rewards being $0.27$. Thus, MSE errors below this level, such as those shown in figure \ref{fig:learning_curves} which reach below levels of $0.01$ indicate the trend investors having developed good and economically-profitable strategies. We prove this empirically by analysing the profit of trend investors over time, as is shown in figure \ref{fig:profit}.

\subsection{Target generation of value investors}

Value investors are initialised with a static target of the market price. We distribute these value investor targets according to a mixture model of two equally-proportioned normal distributions with means separated by a model parameter $\Delta \mu$, which extends the model to consider a market that is split between two expectations of the price. This scenario is of high relevance to the financial market which is often active around binary events, such as elections.

\begin{equation}
    E = \begin{cases}
      \mathcal{N}(100 + \Delta \mu, \sigma) & \text{with probability 0.5} \\
      \mathcal{N}(100 - \Delta \mu, \sigma) & \text{with probability 0.5} \\
      \end{cases}
      \label{eq:normals}
\end{equation}

Equation \ref{eq:normals} outlines the process by which we draw value investor targets. We choose to use a value of $\sigma=5$ arbitrarily based on our typical ranges for $\Delta \mu$, which varies up to 20.

\subsection{User-defined parameters}\label{sec:params}

The model has the following list of parameters that can be determined by the user.

\vspace{0.2cm}
\begin{itemize}

    \item \textbf{\textit{n-value-investors}} (default 10) determines the number of value investor agents that will initialised in the model.
    \item \textbf{\textit{n-trend-investors}} (default 5) determines the number of artificially intelligent trend investor agents that will initialised in the model.
    \item \textbf{\textit{n-dealers}} (default 10) determines the number of market maker agents that will initialised in the model.
    \item \textbf{\textit{bid-offer}} (default 1.0) specifies the fixed interval between each market maker's bid and offer prices.
    \item \textbf{\textit{dealer-position-limit}} (default 20) represents a soft limit to market maker inventory before adjustments are made to prices from their advertised levels. This is the parameter $L$ in equation  \ref{eq:analyticSolution}.
    \item \textbf{\textit{prob-of-link}} (default 50) is the probability of a link forming in our random network as a percentage.
    \item \textbf{\textit{trade-size-cap}} (default 3) represents the maximum trade size a value investor can make. 
    \item \textbf{\textit{market-disparity}} (default 20) determines the difference between the peaks of the bi-modal distributions, from which the value investors targets are drawn, and 100. This is equivalent to the parameter $\Delta \mu$ in  \ref{eq:normals}.
    \item \textbf{enable-broker-market} (default ON) triggers the behavior whereby market makers will transact versus one another if their soft limit is breached. If this parameter is turned off, market makers do not transact against each other.
\end{itemize}
\vspace{0.2cm}

The model also includes the following additional functionalities.

\vspace{0.2cm}
\begin{itemize}

    \item \textbf{Force market makers Short} is a button that instantaneously positions all market makers to be beyond their soft limit. This can be used to demonstrate certain technicals within the market.
    \item \textbf{Market Crash} is a button that instantaneously re-calibrates all value investor targets to be 20\% lower than they were previously.
    \item \textbf{Remove Value Investors} is a button that removes all value investors from the system, even mid-simulation. This allows for training of the RL algorithm in the trend investors before removing them entirely, should the user want to explore these possibilities.
\end{itemize}

\section{Validation of the model}

We examine the effectiveness of the model through a number of statistical tests that are prevalent throughout the literature \citep{kim_yoon_chang_2004, warusawitharana_2016}. Unless specified otherwise, we use the default parameters for the model outlined in \ref{sec:params}, and only consider results from where all trend investors are fully trained. Specifically, we look for quantitative features in the price distribution that reflects the fat-tailed distributions observed in nature, such as a kurtosis above 3 and a Zipfian power-law between price movement sizes and ranks. We additionally illustrate a number of qualitative features exhibited by the model that highlight further similarities to reality.

\begin{figure}
    \centering
    \includegraphics[scale=0.3]{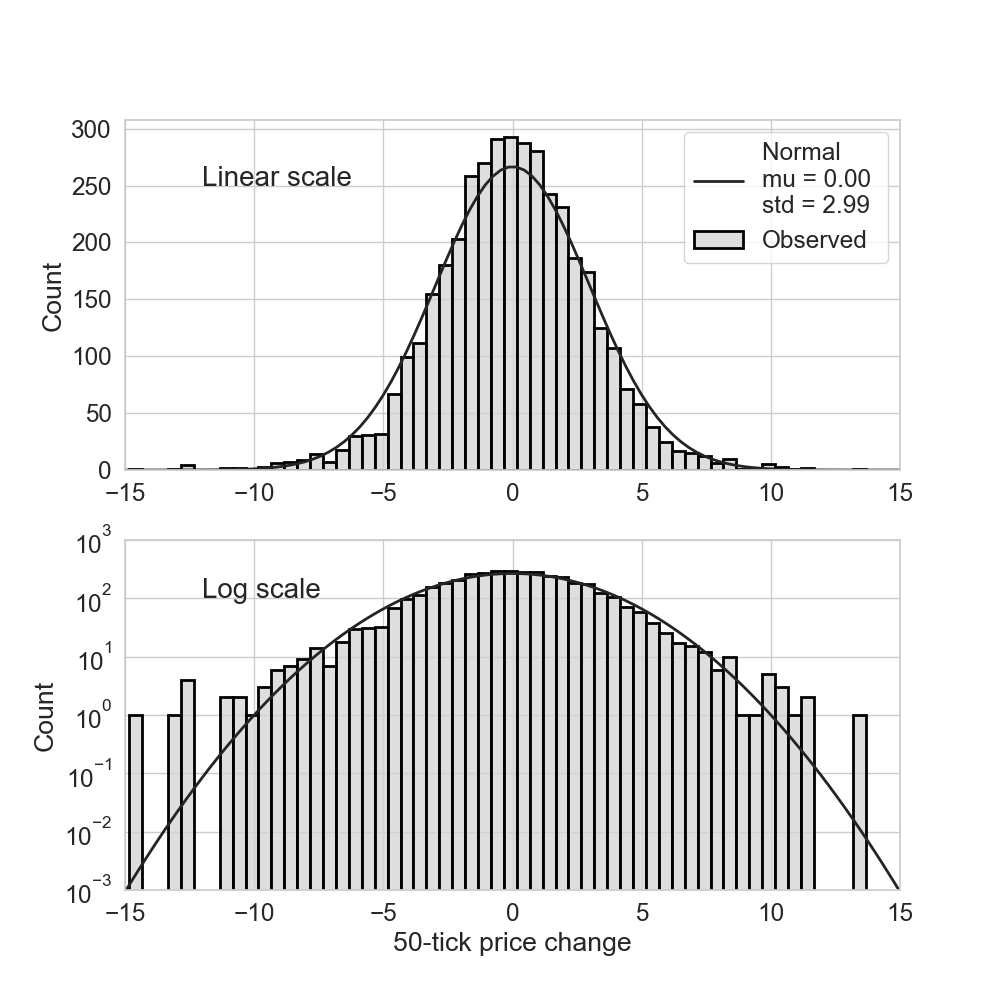}
    \caption{Price changes, measured every 50 iterations of the model once trend investor training has completed, illustrated with a fitted normal distribution. Shown both on a linear scale (upper) and a logarithmic scale (lower).}
    \label{fig:distribution}
\end{figure}

Firstly, the distribution of price changes within the model can be shown to follow fat-tailed distributions, as illustrated in Fig \ref{fig:distribution}.

We can see from Fig \ref{fig:distribution} that the model produces price returns that are inadequate to be described with a normal distribution: a fact that is particularly evident when looking at the tails of the distribution on a logarithmic scale. Both these fatter tails and the clustering of price-changes that are closer to the distribution mean explicate a distribution with a kurtosis greater than 3. Our model consistently produces price distributions with kurtosises above 3. We also note the nature of the distribution tails, which becomes clear in the lower plot of Fig. \ref{fig:distribution}. As the tails deviate from the normal distribution, they become approximately linear in the logarithmic scale, alluding to a power law relationship. This behaviour has been noted to exist in the observed price movements of the Dow Jones \citep{warusawitharana_2016}.

\begin{figure}
    \centering
   \includegraphics[scale=0.3]{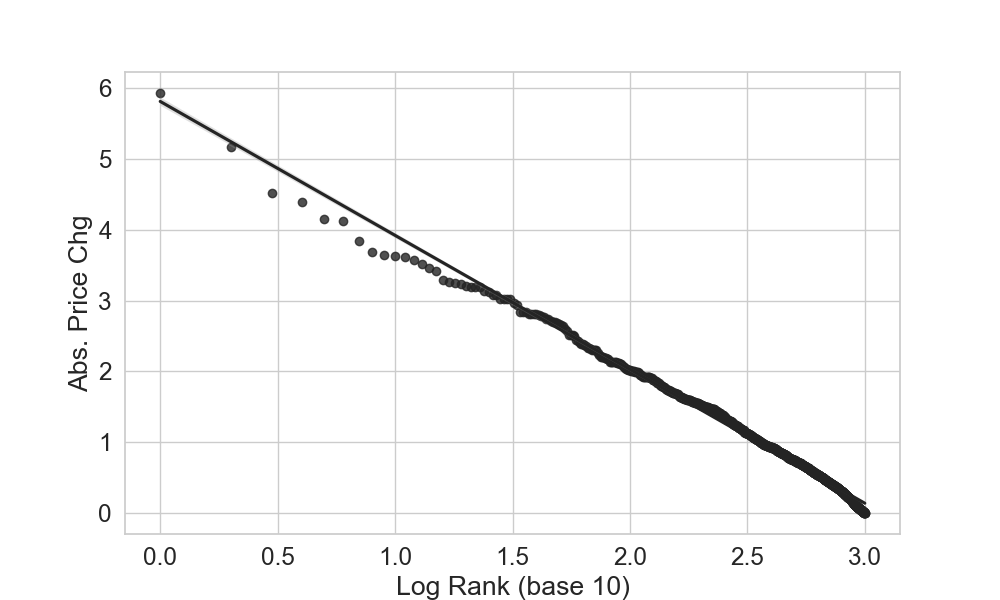}
    \caption{A typical rank distribution of price changes within the model, with changes calculated over 50 time-steps, and illustrated on a logarithmic axis. The linear line implies a power law distribution (Zipf's law).}
    \label{fig:zipf}
\end{figure}

We can directly show that the price changes produced by our model follow a power law, also known as Zipf's law \citep{ZipfOrig}. Although power laws are found throughout nature, particularly in systems comprising multiple agents, this is a comparison that has not been included in the previous literature on agent based models of financial markets. Zipf's law has been demonstrated to apply in practice to stock price movements \citep{kim_yoon_chang_2004}, and it is promising that our model replicates this phenomenon as shown in Fig. \ref{fig:zipf}.

\begin{figure}
    \centering
    \includegraphics[scale=0.3]{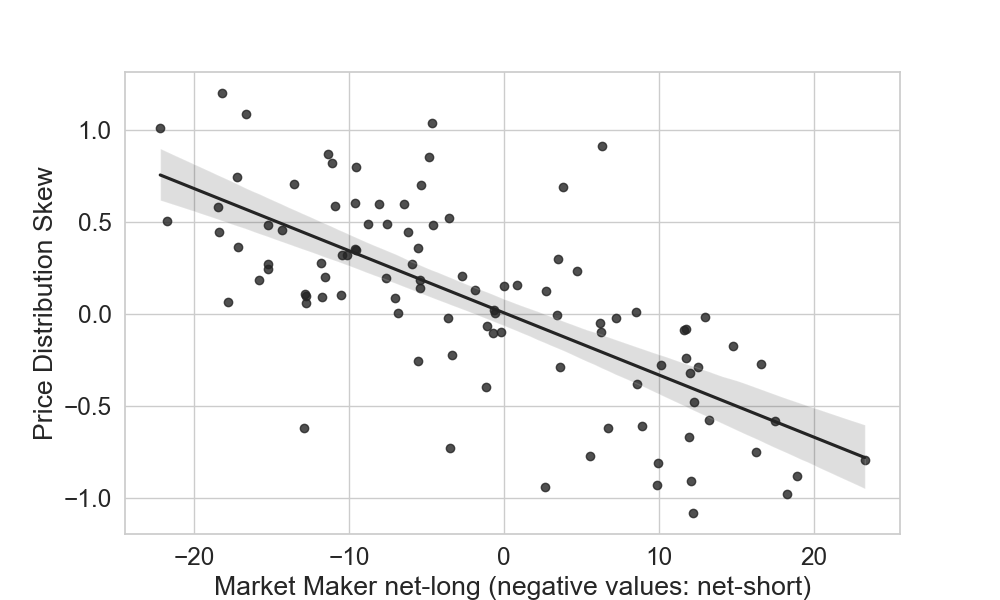}
    \caption{Market maker net positioning (positive equates to long positioning) compared to the skew of the resulting price distribution. The market maker positioning is calculated as an average across the duration of the simulation, and the pricing distribution comprises the cumulative changes over each successive 50 iterations of the model.}
    \label{fig:skew}
\end{figure}

We also find that our model exhibits phenomena that qualitatively match what we would expect in a real financial market. For example, the skew within the price distribution is negatively proportional to the positioning of market makers. Since data on market maker positioning is highly confidential, it is impossible to make quantitative comparisons to data. However, this relationship is highly intuitive, reflects the effect of the model's market makers competing to reduce their inventory, and is a notable feature of the proposed model. This relationship is shown in fig. \ref{fig:skew}.

Further, we note the tendency of the market price to settle at-or-close-to the numerical mean of value investor targets. Again, although this phenomena is difficult to prove in the real world (because value investor target prices are proprietary and confidential), it intuitively exists within real-world markets, with participants routinely interpreting price levels between binary outcomes as an indicator of probability. This feature is a necessary component of the efficient market hypothesis, and in this regard, the market acts as an ensemble model with a mean estimator across its participants. As in real financial markets, our model demonstrates this property for the majority of situations but not all: there are situations where the market price can diverge largely from this mean target and the behaviour is lost, particularly in extreme instances of high volatility.

\begin{figure}
    \centering
    \includegraphics[scale=0.65]{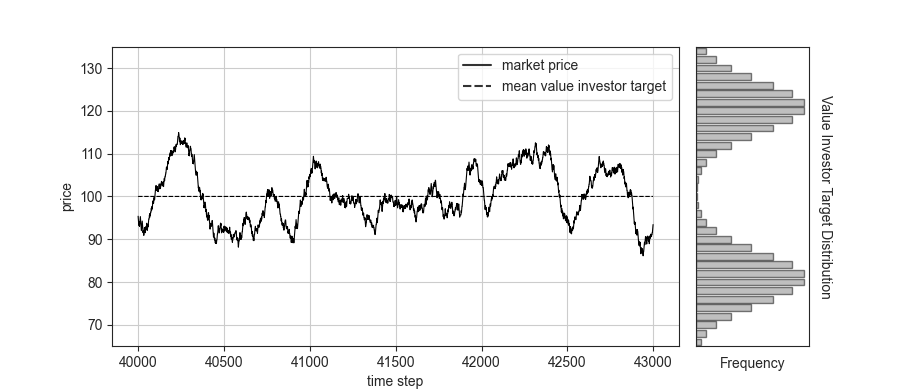}
    \caption{(Left) Price history over 3000 iterations (solid) compared to the mean value investor target (dashed). (Right) Twin-peaked distribution representing value investor targets. To illustrate the effect of price convergence to value investors' mean target in the most generalised setting, we remove the constraint that the bimodal Gaussian distribution has equal probabilities associated with each Gaussian mode.}
    \label{fig:meanDemo}
\end{figure}

The price within the system converges to the mean regardless of our chosen distribution of value investor targets. As is illustrated in Fig. \ref{fig:meanDemo}, the price quickly converges to the mean view in a system where value investor targets are drawn from a bimodal mixture distribution with two peaks. This builds on our previous findings, suggesting that the mean target of value investors acts as an attractor even when the targets are distributed arbitrarily. This behaviour is caused by the value investors scaling their trade size by distance-from-target, as outlined in  \ref{eq:analyticSolution}. The varying trade size with price gives rise to an equilibrium level. We hypothesise that the convergence to the unweighted mean is a direct result of all value investors sharing the same proportionality between difference-to-price-target and trade size. This parameter, $\sigma_{vi}$, is introduced in  \ref{eq:size}.

\begin{figure}
    \centering
    \includegraphics[scale=0.3]{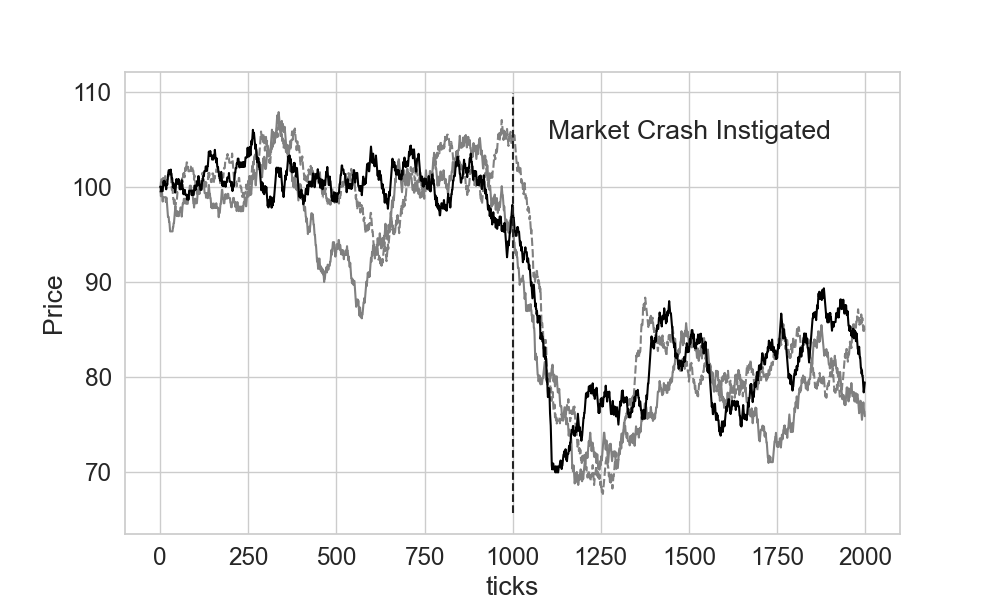}
    \caption{Price history when a market crash is instigated after 1000 iterations of the simulation (vertical dashed) shown for four separate runs. We notice a characteristic overshoot of the initial drop followed a rebound, and increased volatility following the crash compared to before.}
    \label{fig:technicals}
\end{figure}

We also note the existence of patterns and trends within the model's price movements. These are particularly interesting from the perspective of technical analysis (the practice of predicting future prices using only historical price information), which is used by our trend investors. One example is during large price movements, when the market price will typically overshoot the equilibrium level before retracing a smaller amount. This is a widely observed phenomena in the real world that is visible following the vast majority of market crashes. We recover this same technical within our model, as shown in figure \ref{fig:technicals}.

In order to induce the market crash displayed in figure \ref{fig:technicals}, we instantaneously reduce the price targets of all value investors by 20\%. Our prior findings that the market's equilibrium level is strongly influenced by the value investor targets imply that this should be an effective approach to instigating a market crash. Additionally, this is a realistic approach, since sharp drops in financial prices are often prompted in the real world when new information is revealed (via news updates or the publication of financial results) which moves investor expectations. As can be seen in fig. \ref{fig:technicals}, the model exhibits a characteristic rebound following the initial overshoot of the price. The overshoot occurs entirely through a combination of market maker positioning as they overshoot the equilibrium price in their plight to recycle inventory, along with trend investor strategy. This behaviour has been noted in prior ABM literature, which through the empirical study of liquidity-driven crises has concluded that financial crashes exhibit this artifact because of market makers competing to reduce their inventory against the direction of the market \citep{kirilenko2017flash}. Following the initial drop, market makers within our model are eventually able to reduce their inventory during the initial price overshoot, in which value investors actively play a role in purchasing market maker inventory that is being offered at a lower price than their target level. Whilst the market price is finding its new equilibrium level, the scale of price movements is visibly increased, demonstrating the auto-correlation and clustering of volatility. This is also a significant feature present in the real financial markets, with volatility clustering being used as a key parameter in sophisticated models such as the SABR model \citep{hagan2002managing} - one of the most widely used options-pricing models in the modern world.

\section{Reinforcement learning analysis}

Within our model, each trend investor is initialised with their own deep-Q network, therefore allowing dynamically evolving strategies and competition between them and their environment. Whilst the convergence of the models was verified as in section \ref{deepQsec}, we aim here to present an analysis of the strategies being learned by these artificially intelligent agents.

\begin{figure}
    \centering
    \includegraphics[scale=0.5]{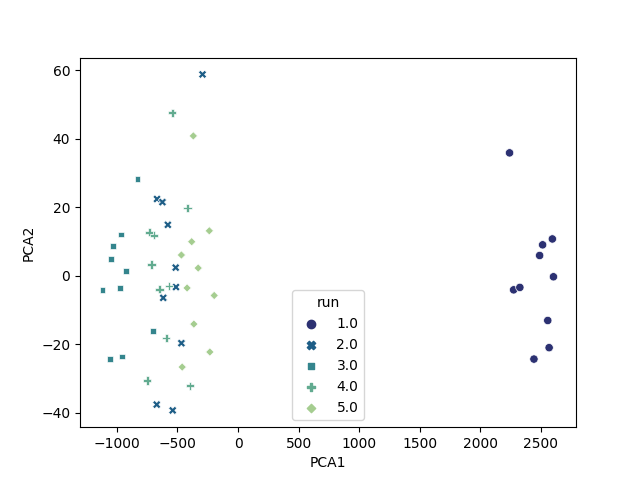}
    \caption{To visualise the trend investor Q-network weights, we have plotted the first two principal components of each trend investor over 5 separate runs. In this simulation, 10 trend investors were initialized, along with the default parameters outlined in section \ref{sec:params}. Whilst agents within the same environment with one another tend to develop similar weights, two distinct groupings of strategies can be seen, separated primarily by the value of the trained weight's first principal component. We hypothesize that the models with similar weights (and therefore similar positions on the chart) also share similar strategies when trading. The snapshots of weights were taken with the default setup presented in section \ref{sec:params}, 100 time-steps after the agent's $\epsilon$ reached its terminal value.}
    \label{fig:PCAweights}
\end{figure}

As can be seen in figure \ref{fig:PCAweights}, two distinct clusters of weight optimums appear. We have visualised these two clusters using principal component analysis. Plotting the principal component of the fully-trained trend investor's weights, we are able to visualise a low-dimensional representation of the weights similarities on a scatter plot.

\begin{figure}
    \centering
    \includegraphics[scale=0.50]{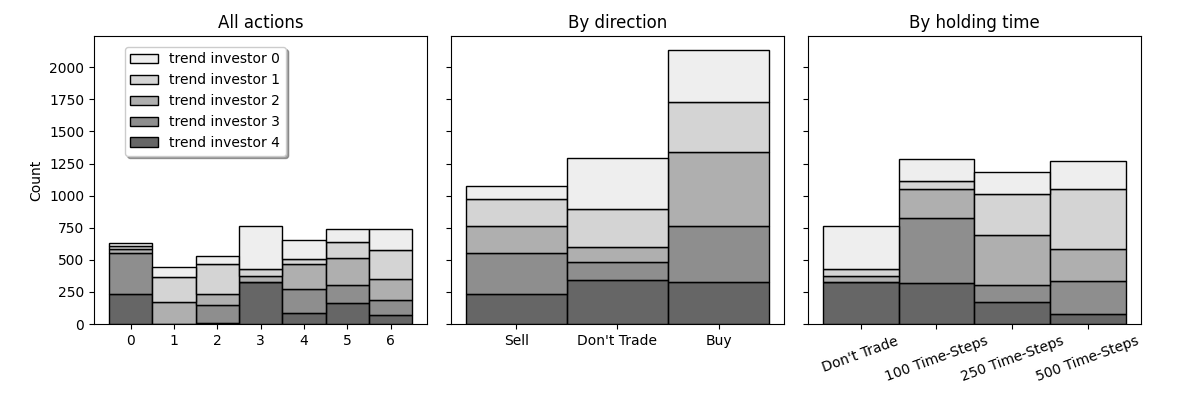}
    \caption{Histograms showing the frequency of different actions choices between each of five agents present in one particular run of the model. In the left-most plot, actions are identified with indexing numbers 0-to-6, respectively representing selling for 100, 250, 500 time steps, doing nothing, and buying for 100, 250 and 500 time steps. Actions were recorded from the point at which all trend investors had completed training (as measured by their $\epsilon$ parameter reaching its minimum value, and recording until each trend investor had taken an additional 1000 actions from that point). We can see a broad spread of actions selected, with this example demonstrating a  preference across the group to buying. Agents exhibit marginal differences in their behaviour, with \textit{trend investor 4} opting for shorter duration trades, whilst by comparison, \textit{trend investor 1} chose longer time horizons for their trades.}
    \label{fig:actionHist}
\end{figure}

\begin{figure}
    \centering
    \includegraphics[scale=0.5]{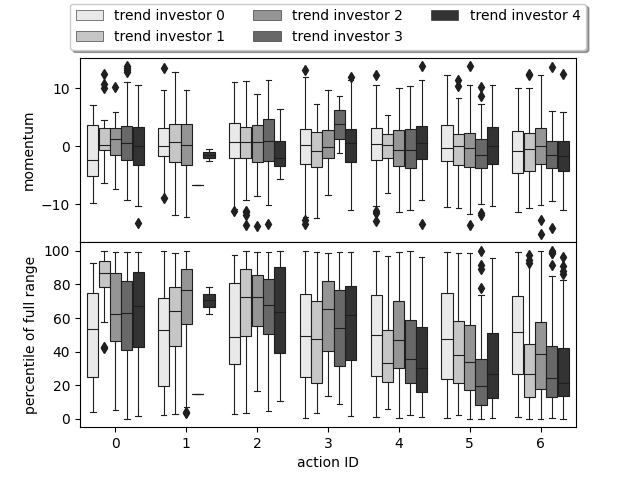}
    \caption{We analyse the state corresponding to each trend investor decision. The action-state pairs are recorded for each trend investor over the 1000 actions that the agent makes after its exploration parameter $\epsilon$ parameter used in training reaches its final value. The action IDs respectively correspond to selling for 100 time steps, 250 time steps, and 500 time steps, doing nothing, and buying for 100, 250 and 500 time steps. We measure two metrics regarding price movements: momentum (upper plot) is defined as the difference between the product's price over the immediately preceding 50 model iterations (where price is the mid price between the most competitive bid and offer prices available to each agent via the network); and we measure mean reversion trading strategies by considering where the product's price is relative to it's full observed trading range, which is represented as a percentile (lower plot).}
    \label{fig:strategies}
\end{figure}

The lower chart in figure \ref{fig:strategies} shows that the trained trend-investors select actions corresponding to selling (which have action IDs of 0, 1 and 2) when the price is at the upper end of it's historical range, and select actions corresponding to buying (which have action IDs of 4, 5 and 6) when the price is at the lower end of its historical range. This is a strong indication that the trend investors are primarily learning a strategy that corresponds to mean-reversion trading. This is the most intuitive strategy for our given set-up, since the deterministic behaviour of value investors has the effect of bounding the price range. The trend investors within our market compete over profit, since each trend investor trade that follows a mean-reversion strategy will implicitly reduce the size of the available profit following the same strategy that is available to the other trend investors. For example, in the act of selling when the price is at the top of its historical range in anticipation of the price returning towards the center of its trading range, the trend investor would transact with a market maker thus placing downward pressure on the market price, and reducing the scale of the price move. This is central to a trend investor's function as a dynamic learner: by continually adapting to profit from the market's movements, the trend investors reduce the size of the market's inefficiencies. We take the empirical evidence that the trend investors develop profitable strategies from figure \ref{fig:profit} as validation that they are acting towards this end.

\section{Investigations into market structure}
\label{sec:networkresults}
By altering the network structure within the model, we are able to analyse the system under different market structures. This is facilitated both through our novel consideration of a network topology, and also our use of market makers as the sole intermediaries within the system. Prior literature that has used an order book or exchange as the intermediary cannot perform these investigations, since with one central intermediator, all agents are only connected through one edge to the central node.

\begin{figure}
    \centering
    \includegraphics[scale=0.6]{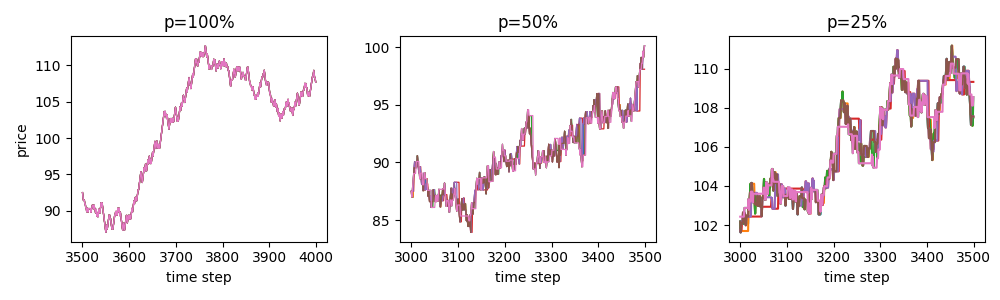}
    \caption{Illustrations of each market maker's mid price across three simulations with values of $p$ equal to 100\% (left), 50\% (middle) and 25\% (right). As the value of $p$ decreases, visibility within the market also decreases, and the disparity between market maker prices increases.}
    \label{fig:histFragmentation}
\end{figure}

We focus on varying the probability of a link forming within our network structure outlined in section \ref{sec:network}, denoted as $p$. Decreasing the value of $p$ has the effect of limiting the visibility of agents within the model. This is an important consideration of an OTC market, which has not been considered in prior work. We observe that in more fragmented network structures, with fewer links and a lower value of $p$, market maker prices begin to diverge from one another. This is illustrated in figure \ref{fig:histFragmentation}. 

\begin{figure}
    \centering
    \includegraphics[scale=0.6]{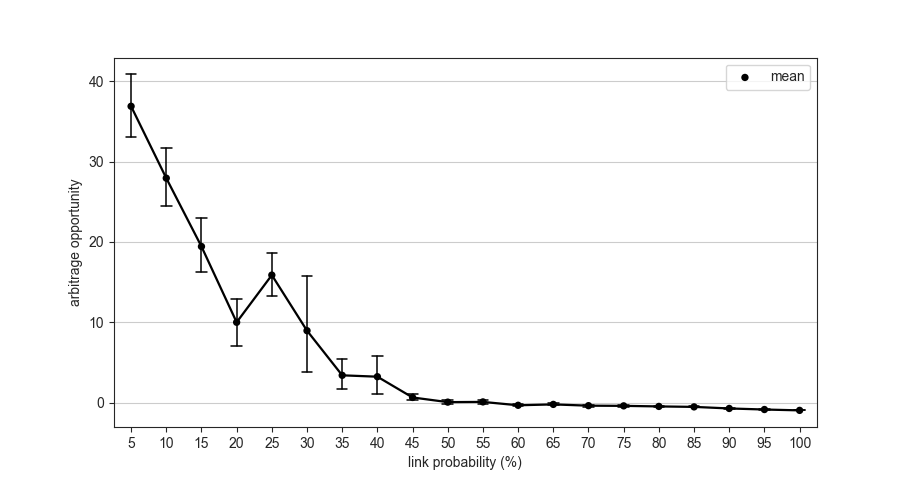}
    \caption{Below a network parameter $p$ of approximately 30\%, arbitrage opportunities between market maker prices rapidly grow in size. Arbitrage opportunity is measured as the price differential between the most competitive bid and the most competitive offer between all market makers. Arbitrage opportunity is positive when an external investor can profit when instantaneously buying from one market maker, and selling to another.}
    \label{fig:arbFragmentation}
\end{figure}

In an efficient market, market makers' prices should differ by a relatively small amount. However, in reality, arbitrage opportunities can emerge when one market maker's buy price drifts above the sell price of another market maker. The arbitrage in this case is defined as the profit that would be made from instantaneously buying from one market maker, and selling to the other. Whilst arbirage is not directly capitalisable by any agents within our model (since the system is limited to one trade per time-step), we can directly observe this arbitrage opportunity. Defining the arbitrage opportunity as the difference in price between the most competitive bid price, and the most competitive offer price across all market makers, we find that as the value for $p$ decreases towards approximately 30\%, arbitrage opportunities increase in frequency. These results are displayed in figure \ref{fig:arbFragmentation}.

\begin{figure}
    \centering
    \includegraphics[scale=0.55]{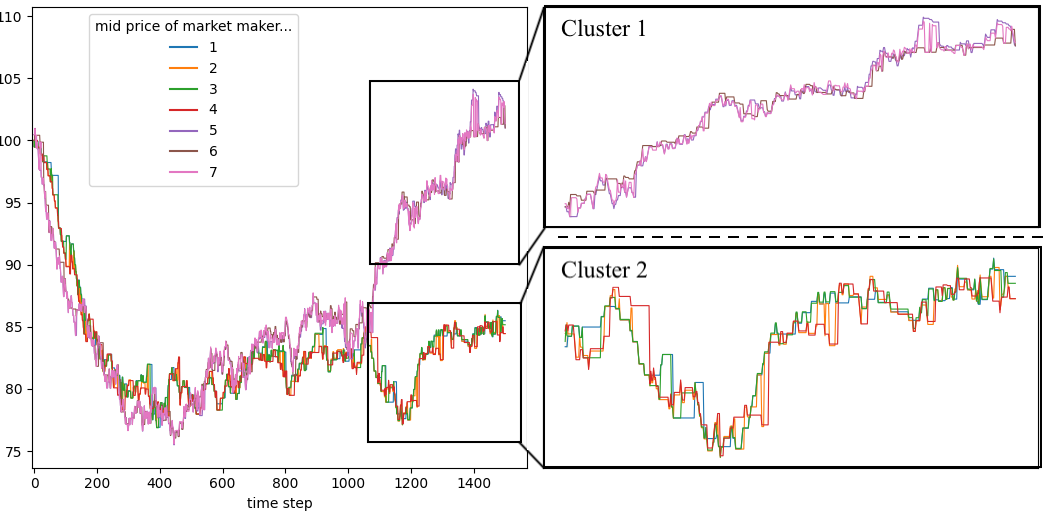}
    \caption{Simulation of the model with link probability $p$ of 8\%, and the default values for the remaining parameters. By plotting all market maker mid-prices, we can see that the market fragments into two smaller clusters at this low value for $p$. Each cluster appears to behave as a smaller, independent market with its own dynamics. The price change distributions of each cluster, when considered individually, are both fat-tailed and are illustrated in figure \ref{fig:fragmentation_net}. The above example illustrates the market fragmenting into two separate components, however, larger numbers of fragments are also possible depending on the initialisation parameters.}
    \label{fig:market_fragmentation}
\end{figure}

\begin{figure}
    \centering
    \includegraphics[scale=0.18]{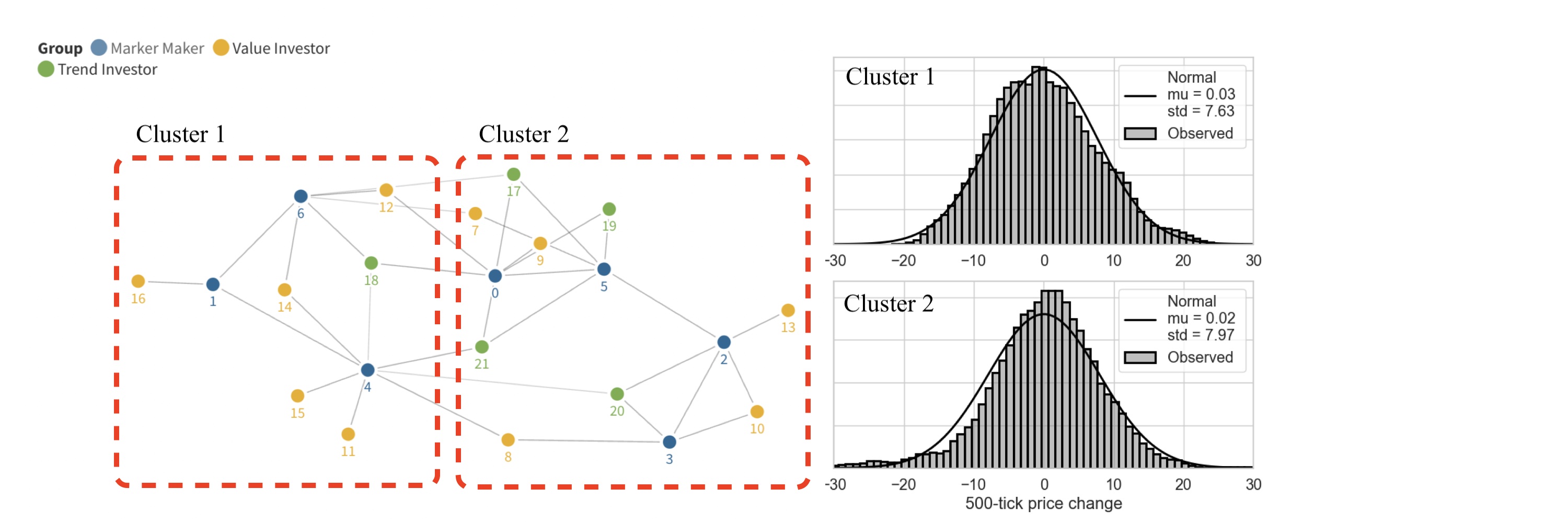}
    \caption{Visualisation of the network topology measured in figure \ref{fig:market_fragmentation}. Each of the two clusters formed in figure \ref{fig:market_fragmentation} is represented by two distinct groups of market makers (blue nodes), which are outlined in red on the left of this figure. Neither cluster has any inter-connected or shared market makers. The price change distributions of each cluster are shown on the right of the illustration, and we notice the fat-tailed nature of both distributions. The price change distribution of cluster 2 in this example is super-normal with a kurtosis 4.32. The price change distribution of cluster one has a kurtosis of 2.85, and is sub-normal. The numbers labelling nodes within the network diagram are identification numbers.}
    \label{fig:fragmentation_net}
\end{figure}

As can be seen in figure \ref{fig:histFragmentation}, when $p$ is greater than 30\%, these arbitrage opportunities typically last for short periods of time, and are relatively small. However, for values of $p$ lower than 30\%, the market undergoes a phase change, where these arbitrage opportunities rapidly grow in size. This phase change is not only characterized by a dramatic increase in arbitrage opportunity, but also by the nature of the market prices that ensue. As shown in figure \ref{fig:market_fragmentation}, this phase change involves a separation of the market into multiple, distinct clusters. The example shown in figure \ref{fig:market_fragmentation} shows an example where the market is split into two clusters. Each cluster behaves as a sub-market within the overall system, with it's own prevailing price, complex behaviour and supply and demand dynamics. We infer that each cluster that forms when the market fragments comprise a group of agents with relatively high inter-connectivity, and there is a minimal, but not necessarily zero, number of edges between different clusters. Figure \ref{fig:fragmentation_net} illustrates the simulation's network topology, which demonstrates that the behaviour seen in figure \ref{fig:market_fragmentation} corresponds with the market markers being separated into two unconnected clusters. Figure \ref{fig:fragmentation_net} also demonstrates that each of the two prices produces a fat-tailed price change distribution, with a kurtosis above 3.

\begin{figure}
    \centering
    \includegraphics[scale=0.6]{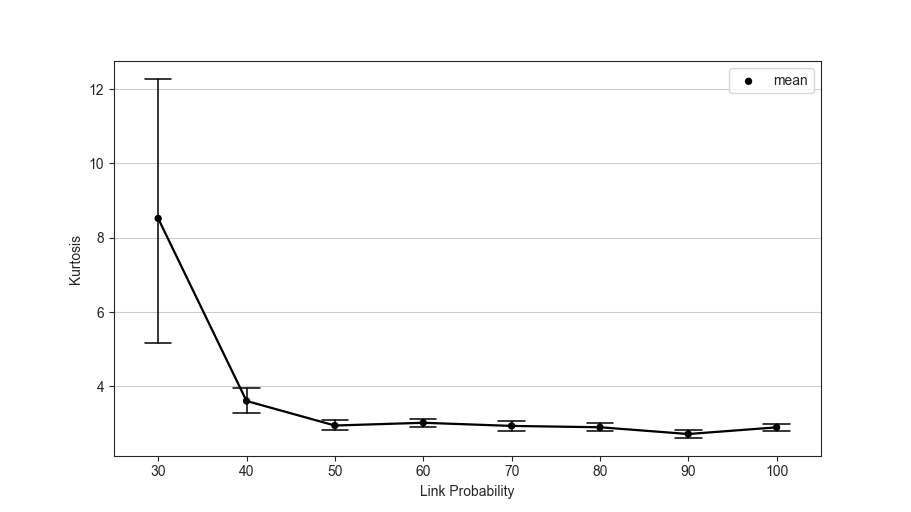}
    \caption{As the link probability $p$ decreases, and the network connecting the market becomes increasingly sparse, the kurtosis of the resulting price change distribution also increases. As discussed, for values of $p$ as high as 30\%, the market can fragment into entirely distinct sub-markets. The results show the averages of 8 simulations for each link probability setting.}
    \label{fig:kurtosisFragmentation}
\end{figure}

The phase change also has an effect on the overall kurtosis of the resulting price change distribution. As shown in figure \ref{fig:kurtosisFragmentation}, we can see that for values of $p$ lower than approximately 40\%, kurtosis steeply rises from values of around 3, to be increasingly fat-tailed. We can conclude from this that limited visibility within financial markets can be a contributing factor to their characteristic fat-tailed distribution. We hypothesise two reasons for this: firstly, that limited visibility impairs the market makers' ability to recycle risk to the investors. As such, it is plausible that a build-up of inventory within the market maker's portfolios could give rise to more frequent dramatic price changes within the market. Secondly, once the market has fragmented into separate clusters (as is relevant for values of $p$ below approximately 30\%), it is plausible that each of the clusters (in isolation) exhibits a price change distribution that differs from the other clusters. Our measure of kurtosis throughout this paper is from a distribution comprised of all market maker price movements. Thus, we are implicitly summing the price change distributions across all of the clusters, which we hypothesise could have an increasing effect on the kurtosis value.

\section{Discussion}

We have presented a novel agent-based model for simulating an OTC financial market, that is unique in its consideration of limited market visibility, and its implementation of market makers as the sole intermediaries. With this approach, combined with a network topology to the underlying market, we are able to perform investigations into the effect of a market's structure on it's function. The models mechanics have been outlined, and demonstrated to replicate a wide range of features that are observed in real financial markets. For example, price changes follow fat-tailed distributions that follow Zipf's law when ranked \citep{kim_yoon_chang_2004}, and a power distribution attenuation in the tails \citep{warusawitharana_2016}. The model also reproduces a number of qualitatively desirable features, such as a negative correlation between market maker positioning and the skew of the price change distribution, a tendency to converge to the mean view of investors, and the exhibition of characteristic and repeatable price patterns. 

We have used our model to examine the effect of network topology on the market function. This is a particularly relevant line of examination in OTC markets which often have lower visibility and liquidity than exchange traded environments. We find that markets that conform to the same network model as we have outlined in section \ref{sec:network} exhibit a critical point of fragmentation, beyond which a phase change occurs. Following such a phase change, market function quickly deteriorates as the opportunity to profit from arbitrage between different market maker prices rapidly grows in size. We have demonstrated that this phase change is often characterised by the market fragmenting into distinct clusters of agents. Each cluster in this paradigm behaves heuristically like an independent market, with its own price, agents, and complex behaviour. Further, we have demonstrated that within this model, the kurtosis of the market's price change distribution increases as market visibility decreases. We demonstrated that this increase occurs at the aforementioned phase change of the market, where the market is characterised by a value of $p$ lower than approximately 30\%.

There is room for further work in investigating the market clusters that form beyond the aforementioned phase change. Whilst each cluster contains its own subset of agents, instances can occur when one sole agent is connected to two distinct clusters. In this case, often this sole agent is not able to trade enough volume to pull the prices of both clusters together. As such, an agent such as a value-investor would effectively act to profit from the arbitrage between the prices of both clusters. However, the influence of these bridging agents does have an effect on the prices of each cluster, and we hypothesise that complex dynamics and correlations can exist between the two prices as a result. This would translate naturally to a physical interpretation, where markets are separated by time zone and geography, and it is both intuitive and common (especially for illiquid products) that prices can differ between the financial markets in Asia and Europe or the United States of America.

Currently, our trend investor implementation is based on deep Q learning, which limits the scope of the trend investors' to a discrete action space, which in part facilitated our analyses of the trend investor strategies. Extending these agents to use reinforcement learning algorithms designed for continuous action spaces (for example, actor-critic \citep{bahdanau2016actor}), would be a natural progression. Market makers could also be adapted to also use a machine learning algorithm, and it is particularly relevant that a market maker's bid-offer spread is not always constant in the real world, but varies with a number of factors including volatility. Artificially intelligent market makers could have the ability to choose their bid offer spread independently, with the goal of maximising their profit in a competitive environment. This is an exciting and new paradigm to consider within the MARL literature.
 
\bibliography{biblio}
\bibliographystyle{plain}


\end{document}